%% file: RJwrapper.tex
\documentclass[a4paper]{report}
\usepackage[utf8]{inputenc}
\usepackage[T1]{fontenc}
\usepackage{RJournal}
\usepackage{amsmath,amssymb,array}
\usepackage{booktabs}

\usepackage{listings}
\usepackage{float}
\usepackage{xcolor}

\begin{document}

\sectionhead{Contributed research article}
\volume{XX}
\volnumber{YY}
\year{20ZZ}
\month{AAAA}

\begin{article}
  \input{RJtemplate}
\end{article}

\end{document}

%% file: RJtemplate.tex
\title{dsld: A Socially Relevant Tool for Teaching Statistics}

\author{by Aditya Mittal, Taha Abdullah, Arjun Ashok, Brandon Zarate Estrada, Shubhada Martha, Billy Ouattara, Jonathan Tran, and Norman Matloff}

\maketitle


\abstract{
The growing influence of data science in statistics education requires tools that make key concepts accessible through real-world applications. We introduce ``Data Science Looks At Discrimination'' (\texttt{dsld}), an R package that provides a comprehensive set of analytical and graphical methods for examining issues of discrimination involving attributes such as race, gender, and age. By positioning fairness analysis as a teaching tool, the package enables instructors to demonstrate confounder effects, model bias, and related topics through applied examples. An accompanying 80-page Quarto book guides students and legal professionals in understanding these principles and applying them to real data. We describe the implementation of the package functions and illustrate their use with examples. Python interfaces are also available.
}


\section{Introduction}
\input{content/introduction}

\section{The dsld Package and Quarto Book}
\input{content/package}

\section{Detecting Discrimination} 
\input{content/detect-background}

\section{Addressing Bias in Predictive Algorithms} 
\input{content/bias}

\section{Python wrappers}
\input{content/dsldPy}

\section{Discussion}
\input{content/discussion}

\bibliography{RJreferences}

\newpage

\address{Aditya Mittal\\
Department of Statistics\\
University of California, Davis\\
USA\\
\email{admittal@ucdavis.edu}}

\address{Taha Abdullah\\
Department of Computer Science\\
University of California, Davis\\
USA\\
\email{tabdullah@ucdavis.edu}}

\address{Arjun Ashok\\
Department of Computer Science\\
University of California, Davis\\
USA\\
\email{aashok@ucdavis.edu}}

\address{Brandon Zarate Estrada\\
Department of Computer Science\\
University of California, Davis\\
USA\\
\email{bzarate@ucdavis.edu}}

\address{Shubhada Martha\\
Department of Computer Science\\
University of California, Davis\\
USA\\
\email{smartha@ucdavis.edu}}

\address{Billy Ouattara\\
Department of Computer Science\\
University of California, Davis\\
USA\\
\email{bouattara@ucdavis.edu}}

\address{Jonathan Tran\\
Department of Computer Science\\
University of California, Davis\\
USA\\
\email{jsttran@ucdavis.edu}}

\address{Norman Matloff\\
Department of Computer Science\\
University of California, Davis\\
USA\\
\email{matloff@cs.ucdavis.edu}}

%% file: content/introduction.tex
Statistics---the class students love to hate! It's hard to think of a
course less popular, yet required by more majors, than statistics.
Recent studies have found a negative student perception of statistical
courses, ranging from undergraduate to graduate level course work
\citep{naidu2014intro, dani2023intro}. Perhaps relabeling as ``data science'' will help a bit; however, the subject is badly in need of
better motivation. To aid this effort, a number of remedies have been
proposed, ranging from the flipped classroom \citep{kovacs} to stories 
involving Disney characters \citep{peters}. The American Statistical 
Association also has some suggestions \citep{gaise}.

Our package, ``Data Science Looks at Discrimination'' (\texttt{dsld}), takes a different approach by appealing to students' awareness of current social issues \citep{bowen2017intro}. The software provides both graphical and analytical/tabular to investigate potential bias related to race, gender, age, and other sensitive attributes. The package focuses on two main areas:

\begin{itemize}

\item \textbf{Detecting discrimination:} Identify and adjust for confounding variables when analyzing potential disparities. For example, is there a gender wage gap when accounting for other factors such as age, occupation, and number of weeks worked?

\item \textbf{Addressing bias in prediction:} Minimize bias in predictive algorithms for decision-making. For example, in loan approval applications, how can the influence of race, whether direct or through proxy variables, be mitigated?

\end{itemize}

The value of the package is greatly enhanced via the use of a tightly integrated open source textbook, written in Quarto \citep{quarto}. The book is not a user manual for the package, but instead provides a detailed treatment of the statistical concepts themselves, illustrated with \texttt{dsld} examples. This package is intended to be useful for teaching, social science research, internal HR analysis, discrimination litigation, and more.  Both parametric and nonparametric regression models are available.

We note other R packages focusing on analysis of discrimination and related issues: {\tt divseg} \citep{divseg} is concerned with urban racial segregation; {\tt genderstat} \citep{genderstat} is ``...an exhaustive tool developed for the R...programming environment, explicitly devised to expedite quantitative evaluations in the field of gender studies;'' {\tt segregation} \citep{segregation} is a tool for the calculation of relationships in two-way contingency tables, including with grouping, with a typical intended use case being analysis of urban racial segregation. Several packages address the issue of fairness in prediction, including \texttt{fairML} \citep{scutari2023fairml}; \texttt{fairmodels} \citep{wisniewski2021fairmodels}; and \texttt{fairness} \citep{fairness}. 

The remainder of this paper is organized as follows. Section 2 introduces the \texttt{dsld} package and accompanying Quarto book. Section 3 covers the detection of discrimination, while Section~4 shows methods to reduce bias in machine learning. Section 5 presents the Python implementations provided by \texttt{dsldPy}. Finally, Section~6 concludes with a discussion.

%% file: content/package.tex
The \texttt{dsld} package was developed in 2023 by seven undergraduates at the University of California, Davis, under the direction of Norman Matloff. It currently includes 24 functions along with an associated open-source textbook. The package is available on CRAN, with the latest version maintained on \href{https://github.com/matloff/dsld}{GitHub}. 
The associated \href{https://htmlpreview.github.io/?https://github.com/matloff/dsldBook/blob/main/_book/index.html}{Quarto notebook} provides additional examples and introduces key statistical principles.  

\textbf{Some basic notation:} We have a response variable $Y$ related to a vector of covariates $X$, and a sensitive variable $S$; the latter may be continuous,  binary or categorical. $Y$ can be continuous or binary, with coding 1 and 0  in the latter case.  In predicting $Y$ in a new case, the predicted value is denoted by $\widehat{Y}$. 








\vspace{-1em}
\subsection{Introducing dsld}

Many \texttt{dsld} functions wrap existing functions from other packages while 
adding functionality specific to discrimination analysis. To make this concrete, 
consider \texttt{dsldLinear()}, which wraps R's \texttt{lm}. The call form is:  

\begin{lstlisting}[language=, breaklines=true, breakatwhitespace=true]
> dsldLinear(data, yName, sName, interactions = FALSE, sComparisonPts = NULL, useSandwich = FALSE) 
\end{lstlisting}

\noindent The arguments are as follows:

\begin{itemize}
\item \texttt{data}:  data frame.
\item \texttt{yName}: the name of the response variable $Y$.
\item \texttt{sName}: the name of the sensitive variable $S$.
\item \texttt{interactions}: if \texttt{TRUE}, include interaction terms with $S$.
\item \texttt{sComparisonPts}: an argument related to interactions, explained further below.
\item \texttt{useSandwich}: if \texttt{TRUE}, uses the sandwich method to address heteroscedasticity \citep{sandwich};  not directly related to discrimination analysis.
\end{itemize}

If {\tt interactions} are set, the model includes interactions
between the sensitive variable and the predictors. Separate linear models are fit for each level of the sensitive variable, which is essentially statistically equivalent. The presence of interaction terms is a key concept in both the package and the accompanying Quarto book. It helps address several important questions: Given the predictors $X$, is there a meaningful difference in the mean of $Y$ across different levels of the sensitive factor $S$? Is this difference consistent across values of $X$?

In the case of interactions, there is no single ``treatment effect'' of the sensitive variable. One cannot speak of \emph{a single} difference in mean wages between men and women, since the difference may vary with factors such as age. The user can specify comparison points at which to evaluate the effects of the different levels of $S$. Estimated differences in conditional means of $Y$ are then reported for the user-specified data points (\texttt{sComparisonPts}).

The package also includes other graphical and analytical functions, including wrappers to existing packages such as {\tt freqparcoord} as well as standalone implementations. These are used both for preliminary exploration of data, and also for visual illustrations of the results found analytically. 

\begin{figure}[h]
    \centering
    \includegraphics[width=0.8\textwidth]{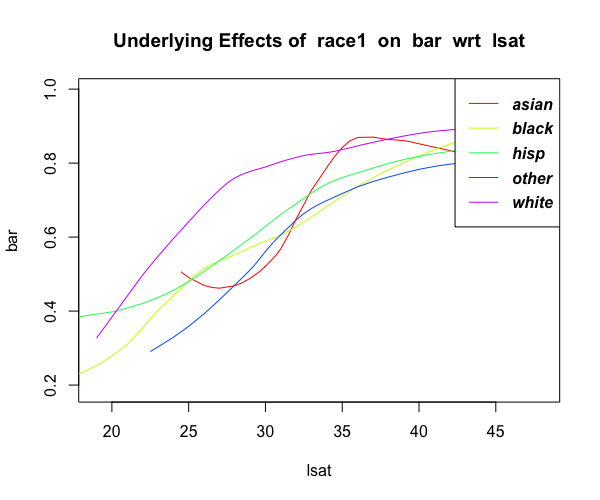}
    \caption{Predicted probability of passing the bar exam by LSAT scores highlighting racial disparities among different groups. \texttt{condits} restricts students with undergraduate GPA below 2.70.}
    \label{fig:conditDisparity}
\end{figure}

For example, one might use \texttt{dsldLogit()} to estimate a logistic model predicting the probability of passing the bar exam based on Law School Admission Test (LSAT) scores, and then supplement the results with a graphical analysis such as Figure~\ref{fig:conditDisparity} using  \texttt{dsldConditDisparity()}. Roughly speaking, there seems to be substantial disparities among the races for low LSAT values, but not as much at the higher end. 

\begin{lstlisting}[language=, breaklines=true, breakatwhitespace=true]
> dsldConditDisparity (data = lsa , yName = "bar", sName = "race1", xName = "lsat", condits = "ugpa <= 2.70" )
\end{lstlisting}

\subsection{The Role of Nonparametric Regression Models}

As noted in the previous section, realistic analyses of discrimination often requires accounting for interaction effects between the covariates $X$ and the sensitive variable $S$. In many cases, it is also necessary to consider nonlinear effects.

Nonlinearity may be addressed with low-degree polynomial models, but modern computing power allows nonparametric regression methods to be applied even on fairly large datasets using a laptop. In \texttt{dsld}, Random Forests (RF) and K-Nearest Neighbors (KNN) are featured as the primary nonparametric approaches \citep{hastie}.

Note that the statistical term \textit{non-parametric regression models}
corresponds in the computer science community with \textit{machine learning} (ML), though there is arguably a difference in interpretation, a running debate ever since statistician Leo Breiman's famous essay, ``Statistical Modeling: The Two Cultures,'' was published \citep{twocultures}.    

We included RFs for a couple of reasons. First, it is arguably the most familiar non-parametric regression method among statisticians, having been developed primarily in the mid-90s and 2000s by statisticians \citep{breiman}. (Less well known, but also significant, is the work by computer scientists such as Tin Ho Kam and her coauthors \citep{ho1995random}.) Second, RFs are accessible to even
nonstatisticians because their ``if–then'' flowchart structure makes them easy to explain and interpret. 

\subsection{The Quarto book}

As mentioned earlier, while the book includes \texttt{dsld} examples, its primary focus is on the underlying statistical concepts rather than serving as a user manual for the package. It places strong 
emphasis on developing a practical understanding of the 
methods, going beyond their formal definitions. Below is a short 
excerpt:

\begin{quote}

One may have specific confounders in mind for a particular analysis, but it is
often unclear as to which to use, or for that matter, why not use them all?... 

Technically, almost any variable is a confounder. The impact may quite
minuscule, but through a long chain of relations among many variables, there
will usually be at least some connection, though again possibly very faint...

...there are several issues to consider not using the full set of variables
i.e.\ every variable other than $Y$ and $S$:

\begin{itemize}

\item It may result in overfitting, resulting in large standard errors.

\item It is unwieldy, difficult to interpret. Many treatments of these issues
speak of a desire for a “parsimonious” model.

\end{itemize} 

\end{quote}

The math involved in the book is minimized, and the material should be
accessible to students who have taken a (noncalculus-based) course in
elementary statistics. (There are a few optional ``starred sections''
covering advanced topics.)   

\subsection{Available Functions}

Below is a table of available functions: graphical and analytical methods for estimation; fair machine learning wrappers for prediction.

\begin{table}[H]
\centering
\renewcommand{\arraystretch}{1.2}
\begin{tabular}{|p{3cm}|p{6cm}|p{4cm}|}
\hline
\textbf{Function} & \textbf{Description} & \textbf{Use Case / Package} \\
\hline
\multicolumn{3}{|c|}{\textbf{Graphical Functions}} \\
\hline
\texttt{dsldFreqPCoord} & Frequency-based parallel coordinates & Visualizing multivariate relationships \\
\texttt{dsldScatterPlot3D} & 3D scatter plots with color coding & Exploring 3D data relationships \\
\texttt{dsldConditDisparity} & Conditional disparity plots & Detecting Simpson's Paradox \\
\texttt{dsldDensityByS} & Density plots by sensitive variable & Comparing distributions across groups \\
\texttt{dsldConfounders} & Confounder analysis & Identifying confounding variables \\
\texttt{dsldIamb} & Constraint-based structure learning & Fitting a causal model to data \\
\hline
\multicolumn{3}{|c|}{\textbf{Analytical Functions}} \\
\hline
\texttt{dsldLinear} & Linear regression with sensitive group comparisons & Regression outcome analysis \\
\texttt{dsldLogit} & Logistic regression with sensitive group comparisons & Binary outcome analysis \\
\texttt{dsldML} & Machine learning with sensitive group comparisons & Non-parametric models (KNN, RF) \\
\texttt{dsldTakeALookAround} & Feature set evaluation & Assessing prediction fairness \\
\texttt{dsldCHunting} & Confounder hunting & Finding variables that predict both Y and S \\
\texttt{dsldOHunting} & Proxy hunting & Identifying variables that predict S \\
\texttt{dsldMatchedAte} & Causal inference via matching & Estimating treatment effects \\
\texttt{dsldFrequencyByS} & Assessment of confounding between $S$ and $Y$ & Frequency estimation across sensitive groups \\
\hline
\multicolumn{3}{|c|}{\textbf{Fair Machine Learning Functions}} \\
\hline
\texttt{dsldFairML} & Wrappers for FairML algorithms & 
\texttt{dsldFrrm}, \texttt{dsldFgrrm}, \texttt{dsldZlm}, \texttt{dsldNclm}, \texttt{dsldZlrm} \\
\hline
\texttt{dsldQeFairML} & Wrappers for EDFFair algorithms & 
\texttt{dsldQeFairKNN}, \texttt{dsldQeFairRF}, \texttt{dsldQeFairRidgeLin}, \texttt{dsldQeFairRidgeLog} \\
\hline
\texttt{dsldFairUtils} & Cross-validation and grid search utilities & Exploring fairness--accuracy tradeoffs \\
\hline
\end{tabular}
\caption{List of \texttt{dsld} functions grouped by category.}
\label{tab:dsldfunctions}
\end{table}

\subsection{Datasets}

Both through the \texttt{dsld} package itself and through the packages it wraps, a number of built-in datasets are available, greatly improving its usefulness of the package as a teaching tool. These datasets are primarily observational in nature, an issue discussed in the \texttt{Quarto} book.

%% file: content/detect-background.tex
Discrimination is a significant social issue in the United States and
in many other countries. There is lots of available data with which one
might investigate possible discrimination, but how might such
investigations be conducted? There is a rich array of classical
parametric methods, and recently nonparametric regression
methods are in increasing use, such as in HR management
\citep{FrissenFair2022}. The \texttt{dsld} package offers both graphical
and analytical tools to detect potential biases, both of which are particularly beneficial for students. These tools help students build intuition and understanding by linking statistical concepts to broader social contexts.

In this section, we use the {\tt lsa} dataset on law school
admissions, focusing on \emph{race} as the sensitive variable. The
racial categories included are Asian, Black, White, Hispanic, and Other.
This dataset is available through the \texttt{dsld} package for further
exploration and analysis.

\vspace{-1em}
\subsection{Graphical/Tabular Methods}

Effective graphs and visualizations can greatly enhance students' 
understanding of data. The \texttt{dsld} package provides a variety of 
graphical methods as shown in Table~\ref{tab:dsldfunctions}.  

As an example, consider potential discrimination in college and graduate school admissions. Standardized testing has faced increasing criticism for disproportionately favoring students with greater family income and resources \citep{foreiter}. Studies have documented differences in test results between Black and White students 
\citep{Dixon2013}. As result, many institutions have removed 
standardized test requirements such as the SAT and GRE from their 
application processes. This debate motivates the following examples to examine biases and potential confounding variables in the \texttt{lsa} dataset.

\vspace{0.5em}
\noindent 
\textbf{dsldConditDisparity}

Suppose we want to analyze the effect of LSAT scores on bar exam passage rates to investigate potential racial disparities. We can compare probabilities of passing at given LSAT scores across racial groups. The function \texttt{dsldConditDisparity()} is well suited for this purposes, as it plots conditional disparities and highlights how outcomes vary by race.

\begin{lstlisting}[language=, breaklines=true, breakatwhitespace=true]
> dsldConditDisparity(data = lsa, yName = "bar", sName = "race1", xName = "lsat", condits = NULL) 
\end{lstlisting}

\begin{figure}[H]
    \centering
    \includegraphics[width=0.8\textwidth]{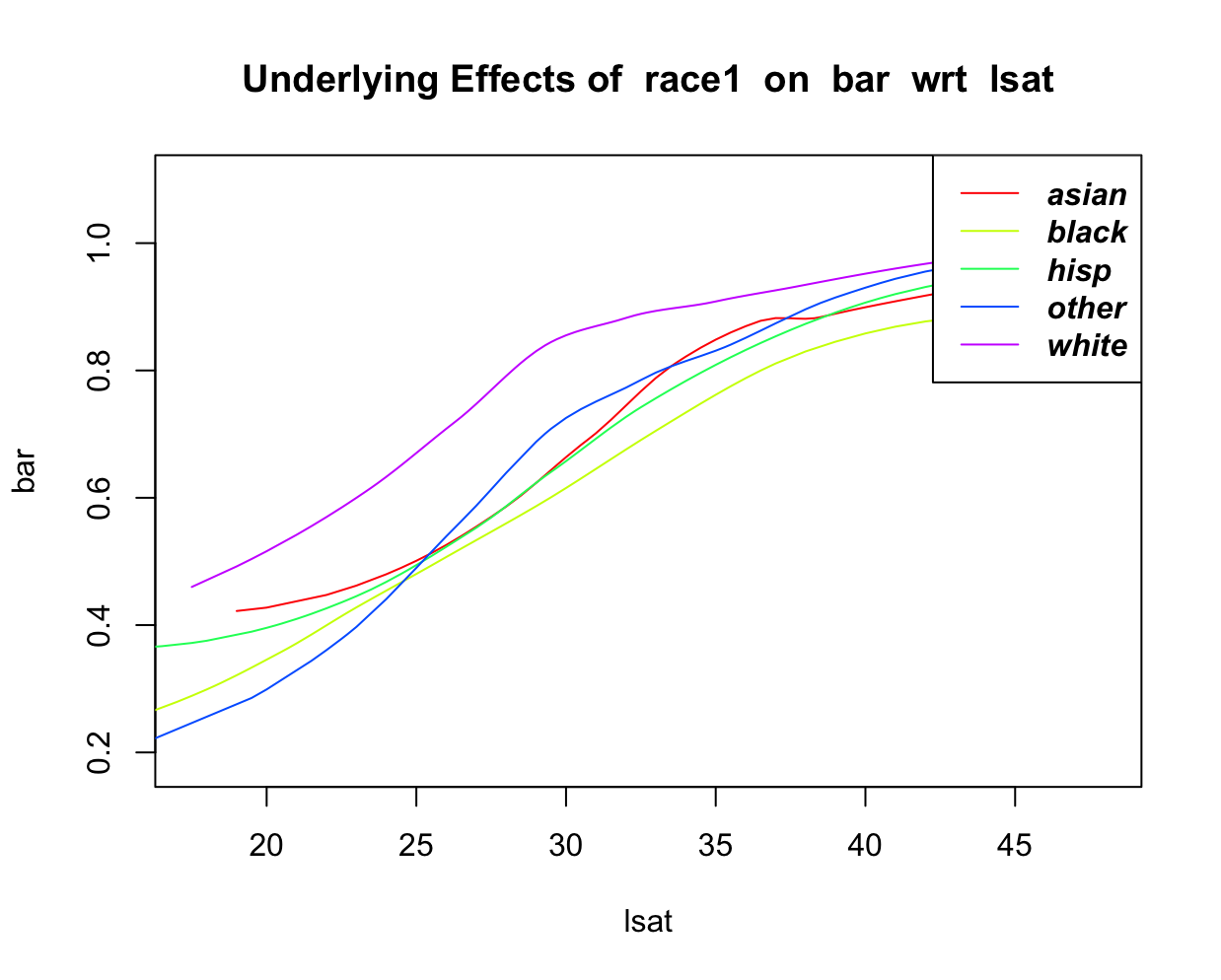}
    \caption{Estimated probability of passing the bar exam by LSAT 
    scores, highlighting racial disparities among different 
    groups. All students.}
    \label{fig:conditDisparity_all}
\end{figure}

Figure~\ref{fig:conditDisparity_all} shows that non-White groups exhibit similar outcomes, with White students having higher passing probabilities at mid-range LSAT levels. At higher LSAT values, passing probabilities are approximately equal across groups, suggesting the relevance of racial interaction terms in analyses and providing a useful basis for classroom discussion. The function includes an optional argument \texttt{condits} (``conditions''). For example, restricting to students with lower undergraduate GPAs ($\leq 2.70$) shows the results in Figure~\ref{fig:conditDisparity}.

\vspace{0.5em}
\noindent 
\textbf{dsldDensityByS/dsldConfounders}

To further examine the relationship between race and LSAT scores, we can use \texttt{dsldDensityByS()} to generate density plots. This function plots the distribution of a response variable $Y$ segmented by a sensitive variable $S$, with optional bandwidth control. In this case, it shows how LSAT score distributions differ across racial groups.

This complements Figure~\ref{fig:conditDisparity_all}, which showed that White students have higher bar exam passage rates than non-White students at the same LSAT levels. Such disparities may be amplified if White students also score higher on the LSAT. To investigate further:

\begin{lstlisting}[language=, breaklines=true, breakatwhitespace=true]
> dsldDensityByS(data = lsa, cName = `lsat', sName = `race1')  
\end{lstlisting}

\begin{figure}[h]
    \centering
    \includegraphics[width=0.8\textwidth]{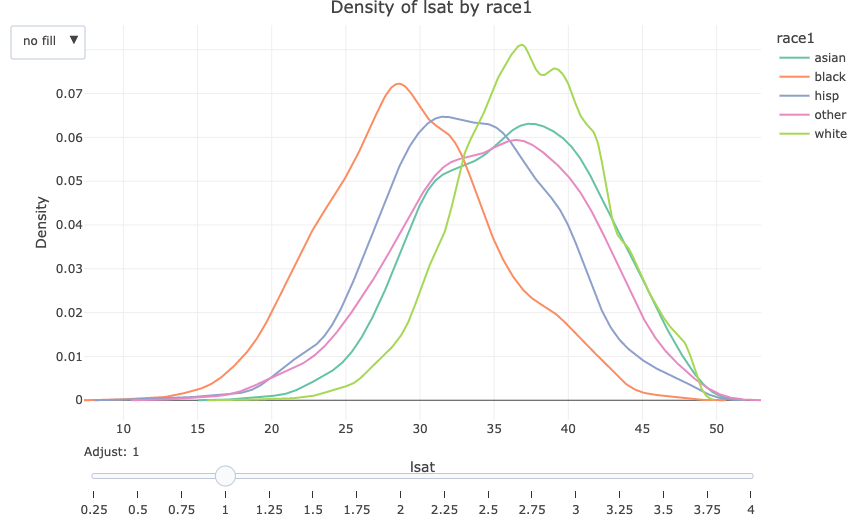}
    \caption{Density plot of LSAT scores segmented by race. The bandwidth parameter can be thought as analogous to controlling the bin width in a histogram to adjust the figure granularity. Default bandwidth is set at 1.}
    \label{fig:lawdensitylsat}
\end{figure}

\noindent 

Figure~\ref{fig:lawdensitylsat} shows the distribution of LSAT scores 
across racial groups, suggesting potential bias. However, such effects 
may be influenced by confounding variables, which affect both dependent 
and independent variables and can lead to spurious associations 
\citep{McNamee2003}. The \texttt{dsldConfounders()} function offers a similar analysis in Plotly for interactive use. For categorical variables, it displays the frequency of the sensitive variable at each level.

\vspace{0.5em}
\noindent 
\textbf{dsldScatterPlot3D}

Further analysis of the relationships among LSAT scores, undergraduate GPA, and family income can reveal potential confounding effects. These 
relationships can be visualized by race using \texttt{dsldScatterPlot3D()}, which provides a 3D perspective on correlations and disparities across groups with respect to income and GPA.

\begin{lstlisting}[language=, breaklines=true, breakatwhitespace=true]
> dsldScatterPlot3D(data = lsa, yNames = c("lsat", "fam_inc", "ugpa"), sName = "race1", pointSize = 4) \end{lstlisting}

\begin{figure}[h]
    \centering
    \includegraphics[width=0.75\textwidth]{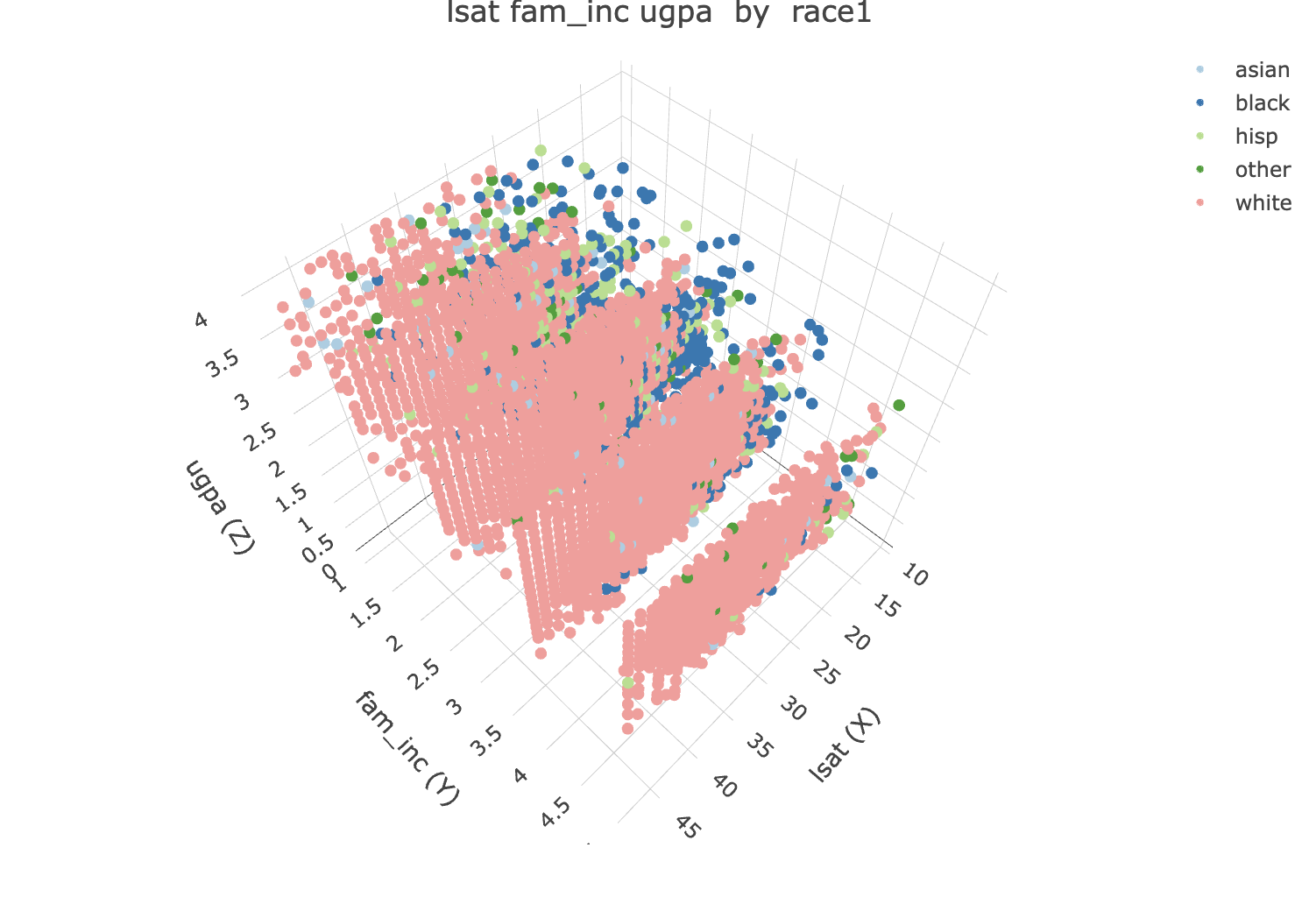}
    \caption{3D scatterplot showing family income, race, and gender. Lower family income is mostly Black and Latino students; higher income levels are predominantly White. Lower LSAT scores are mostly non-white across all income levels. UGPA trends similarly to LSAT but less strongly.}
    \label{fig:scatterplot}
\end{figure}

While only exploratory, Figure~\ref{fig:scatterplot} suggests that family income may \textit{not} be the primary confounder in the relationship between race and LSAT scores. Non-White students appear to consistently score lower across income levels, motivating further discussion of how race influences exam outcomes in relation with other variables. The plot from \texttt{dsldScatterPlot3D()} can be rotated in R’s interactive mode for better visualization.

\vspace{0.5em}
\noindent 
\textbf{dsldFreqParCoord}

The \texttt{dsldFreqParCoord} function plots such interactions using \emph{parallel coordinates} \citep{inselberg}. In a dataset with $p$ columns, each column corresponds to a vertical axis. For every data row, a polygonal line is drawn across the axes, with its height at axis $i$ representing the value of the $i$th variable for that row.

\begin{figure}[H]
    \centering
    \includegraphics[width=0.8\textwidth]{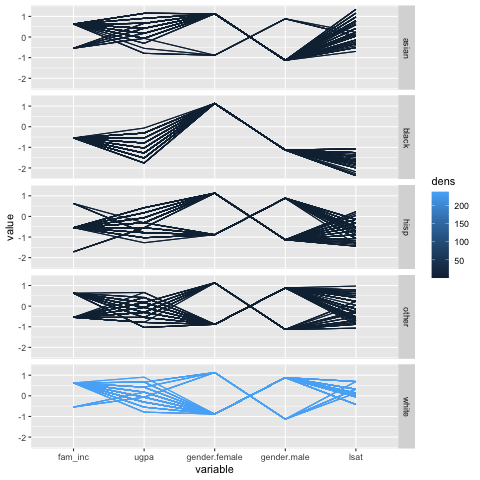}
    \caption{Visualization of the most frequently occurring patterns across different racial groups using parallel coordinates.}
    \label{freqparcoord}
\end{figure}

Each variable is typically centered and scaled, and each data row generates a distinct pattern. The function \texttt{dsldFreqParCoord()} visualizes the $m$ most frequently occurring patterns for each level of the sensitive variable $S$. Two patterns are treated as equivalent if they are proximate in a $k$-nearest neighbor sense, where $k$ is specified by the user. The result is shown in Figure~\ref{freqparcoord}.

\begin{lstlisting}[language=, breaklines=true, breakatwhitespace=true]
> lsa1 <- lsa[,c("fam_inc","ugpa","gender","lsat","race1")] # subset
> dsldFreqPCoord(data = lsa1, m = 75, sName = "race1")
\end{lstlisting}

Several notable patterns emerge. Black students show the least variation, with the most common profile being female students from slightly below-average income families, low undergraduate grades, and very low LSAT scores. Patterns for White students are largely the opposite with greater gender diversity. Asian students show trends similar to Whites, while Hispanic students span a wide range of income levels and grades but display consistently low LSAT scores.

The graphical analysis tools in \texttt{dsld} support teachers by helping visualize complex statistical relationships in datasets.  For example, plots of LSAT score distributions by race can prompt discussion of fairness in standardized testing, and analyses by family income may highlight socioeconomic disparities. These visual methods foster critical thinking by encouraging students to interpret real-world data and consider the societal implications of exploratory and statistical findings. 


\vspace{-1em}
\subsection{Analytical} 

While graphical approaches provide useful exploratory insights, they should be followed by formal statistical methods. The \texttt{dsld} package includes analytical tools for such investigations. Continuing with the LSAT example, suppose we examine pairwise differences in mean LSAT scores across racial groups using a linear model. We illustrate how \texttt{dsldLinear()} facilitates this analysis. This can be extended to the categorical case via \texttt{dsldLogit()}, and to nonparametric comparisons using \texttt{dsldML()}.

\vspace{0.5em}
\noindent
\textbf{dsldLinear (dsldLogit/dsldML)}

A key issue is whether to include interactions between the covariates $X$ and the sensitive variable $S$, a central focus of the \texttt{dsld} package. In educational settings, it is important for students to recognize that in many cases there is no single group difference; instead, the difference may vary with different $X$. This was illustrated in Figure~\ref{fig:conditDisparity_all}, where $Y$ indicated bar exam passage.

We now fit a linear model with $Y$ set to LSAT scores to examine potential bias in the test. We include two versions of the model: one without interaction terms and one with interaction terms.  

\begin{itemize}

\item \textbf{With interactions.}  
In this case, sensitive group effects vary with $X$. The user specifies comparison points through the \texttt{sComparisonPts} argument:

\begin{lstlisting}[language=, breaklines=true, breakatwhitespace=true]
> newData = lsa[c(1,10,100,1000,10000),] 
> z2 = dsldLinear(data = lsa, yName = "lsat", sName = "race1",
+ interactions = TRUE, sComparisonPts = newData)  
> summary(z2) 
\end{lstlisting}

\item \textbf{No interactions.}  
Here, sensitive group effects are assumed constant across $X$:

\begin{lstlisting}[language=, breaklines=true, breakatwhitespace=true]
> z1 = dsldLinear(data = lsa, yName = "lsat", sName = "race1", interactions = FALSE) 
> summary(z1) 
\end{lstlisting}

\end{itemize}

\noindent With interactions, group differences depend on $X$, and the function reports mean differences at user-specified values of \texttt{sComparisonPts}. In general, a linear model with interactions involving a categorical variable with $m$ levels is equivalent to fitting $m$ separate linear models. In the LSAT example, this corresponds to fitting five models, one for each racial group.

Without interactions, a single model is fit, producing one estimated mean difference between pairwise sensitive groups (e.g., Black vs.\ White), independent of $X$. Below we show example from the no-interaction case.\footnote{An example with interactions is provided in the Python section.}

\begin{lstlisting}
$`Summary Coefficients`
     Covariate    Estimate StandardError       PValue
1  (Intercept) 31.98578856   0.448435264 0.000000e+00
2          age  0.02082458   0.005841758 3.641634e-04
3      decile1  0.12754812   0.020946536 1.134602e-09
4      decile3  0.21495015   0.020918737 0.000000e+00
5      fam_inc  0.30085804   0.035953051 0.000000e+00
6         ugpa -0.27817274   0.080430542 5.430993e-04
7   gendermale  0.51377385   0.060037102 0.000000e+00
8   race1black -4.74826307   0.198088318 0.000000e+00
9    race1hisp -2.00145969   0.203504412 0.000000e+00
10  race1other -0.86803104   0.262528590 9.449471e-04
11  race1white  1.24708760   0.154627086 6.661338e-16

$`Sensitive Factor Level Comparisons`
   Factors Compared Estimates Standard Errors      P-Value
1     asian - black  4.748263       0.1980883 0.000000e+00
2      asian - hisp  2.001460       0.2035044 0.000000e+00
3     asian - other  0.868031       0.2625286 9.449471e-04
4     asian - white -1.247088       0.1546271 6.661338e-16
5      black - hisp -2.746803       0.1863750 0.000000e+00
6     black - other -3.880232       0.2515488 0.000000e+00
7     black - white -5.995351       0.1409991 0.000000e+00
8      hisp - other -1.133429       0.2562971 9.764506e-06
9      hisp - white -3.248547       0.1457509 0.000000e+00
10    other - white -2.115119       0.2194472 0.000000e+00
\end{lstlisting}

In addition to the model's coefficients, "Sensitive Factor Level Comparisons" shows mean differences in LSAT scores across pairwise sensitive groups. For example, the estimated Black–White gap is 6.0 points with a standard error of 0.141, indicating a substantial disparity in LSAT scores.

By contrast, the effect of family income is small, providing a useful 
``teachable moment.'' A key concept in statistics is the distinction between \textit{statistical significance} and \textit{practical significance}. Family income, measured in quintiles (1--5), has a coefficient that is highly statistically significant (p-value $\approx 0$), yet the effect size is only about 0.3. The difference between the 3rd and 4th quintiles is therefore just 0.3 points in mean LSAT score---minuscule compared to the overall range of 11 to 48 and the 6-point Black--White gap.

Users may also get additional information such as model coefficients with \texttt{coef()}, covariance matrices with \texttt{vcov()}, and predictions at new data points with \texttt{predict()}. In the interactions case, this corresponds to $m$ separate results, whereas in the no-interactions case it produces a single model as shown above.


\vspace{0.5em}
\noindent 
\textbf{Additional Functions}

\noindent \texttt{dsld} includes additional analytical functions for dataset investigation:  

\begin{itemize}

\item \textbf{dsldCHunting}  
\begin{lstlisting}[language=, breaklines=true, breakatwhitespace=true]
> dsldCHunting(data, yName, sName, intersectDepth = 10)
\end{lstlisting}  
Uses random forests to identify important predictors of $Y$ (without $S$) and $S$ (without $Y$). Reports intersections of top predictors up to \texttt{intersectDepth} to detect potential confounders.  

\item \textbf{dsldOHunting}  
\begin{lstlisting}[language=, breaklines=true, breakatwhitespace=true]
> dsldOHunting(data, yName, sName)
\end{lstlisting}  
Converts factors to dummies and computes Kendall Tau correlations between $S$ and potential proxy variables $O$ (all columns other than $Y$ and $S$). Detect indirect effects of $S$ through $O$, useful in fair machine learning applications.  

\item \textbf{dsldTakeALookAround}  
\begin{lstlisting}[language=, breaklines=true, breakatwhitespace=true]
> dsldTakeALookAround(data, yName, sName, maxFeatureSetSize = (ncol(data)-2), holdout = floor(min(1000,0.1*nrow(data))))
\end{lstlisting}  
Evaluates feature sets for predicting $Y$ while considering fairness–utility tradeoffs. Compares accuracy with and without $S$ and measures predictability of $S$ from $X$.  

\item \textbf{dsldFrequencyByS}  
\begin{lstlisting}[language=, breaklines=true, breakatwhitespace=true]
> dsldFrequencyByS(data, cName, sName)
\end{lstlisting}  
Provides an informal check of whether a variable $C$ may act as a confounder between $S$ and $Y$ by examining frequencies across $S$.  

\end{itemize}

\vspace{-1.5em}
\subsection{Causal Models}   

\noindent Given the importance of causal inference in statistics courses, \texttt{dsld} provides two functions for this purpose: \texttt{dsldMatchedATE()} (analytical) and \texttt{dsldIamb()} (graphical).    

The function \texttt{dsldMatchedATE()}, wrapping \texttt{Matching::Match}, conducts matched-pairs analysis \citep{huber}. Direct pairing can be done, or one may opt to use propensity scores, using either \texttt{glm} or \texttt{qeML::KNN} for a nonparametric  k-Nearest Neighbors approach. 

Let's estimate the ``treatment effect'' of being female 
in a possible gender wage gap. We use the \texttt{svcensus} data, which includes information on individuals' age, education, occupation (mainly engineering fields), income, weeks worked. 

\begin{lstlisting}[language=, breaklines=true, breakatwhitespace=true]
> data(svcensus)
> summary(dsldMatchedATE(svcensus,'wageinc','gender','male'))

# Estimate...  9634.5
# SE.........  380.03
# T-stat.....  25.352
# p.val......  < 2.22e-16
# Original number of observations..............  20090
# Original number of treated obs...............  15182
# Matched number of observations...............  20090
# Matched number of observations  (unweighted).  20090
\end{lstlisting}

\noindent 
Men are estimated to earn \$9634.50 more than similar women, with a standard error of \$380.03. 

Using either logit or k-NN to predict gender, the estimated wage gap becomes:

\begin{lstlisting}[language=, breaklines=true, breakatwhitespace=true]
> summary(dsldMatchedATE(svcensus,"wageinc","gender","male",
   propensFtn="glm"))
   
# Estimate...  10332 
# SE.........  408.39 
# ...

> summary(dsldMatchedATE(svcensus,"wageinc","gender","male",
   propensFtn="knn",k=50))

# Estimate...  9877.8 
# SE.........  439.73 
# ...

\end{lstlisting}    

The function \texttt{dsldIamb} addresses \emph{causal discovery}. Typically, a causal \emph{Directed Acyclic Graph} (DAG) reflects the analyst's intuition about relationships among variables. A DAG cannot generally be derived from the data itself without strong assumptions \citep{shalizi, scutari2023fairml}. 

The \texttt{bnlearn} package provides several such models. \texttt{dsldIamb()} wraps the \texttt{iamb} function, allowing users to construct a basic DAG directly from data.

\begin{lstlisting}[language=, breaklines=true, breakatwhitespace=true]
> svcensus$wkswrkd <- as.numeric(svcensus$wkswrkd) # numeric data
> svcensus$wageinc <- as.numeric(svcensus$wageinc)
> iambOut <- dsldIamb(svcensus)
> plot(iambOut)
\end{lstlisting}

\vspace{-3em}
\begin{figure}[H]
    \centering
    \includegraphics[width=0.71\textwidth]{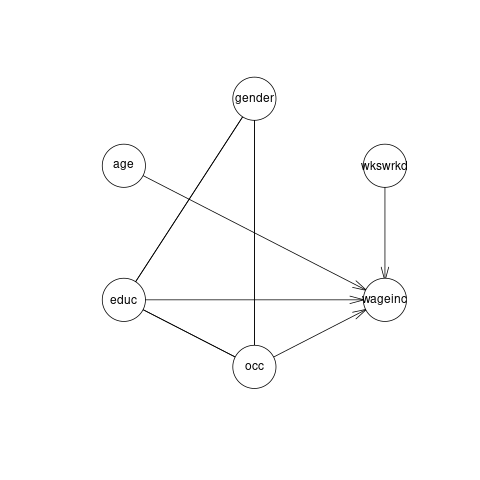}
    \caption{DAG generated under iamb assumptions}
    \label{fig:iamb}
\end{figure}

Figure \ref{fig:iamb} shows the result. Interestingly, the only variable not modeled as causal for wage income is gender, though gender is shown to have a weak, non-directional association with occupation and education. As noted, the graph relies on very restrictive assumptions and should be regarded as exploratory rather than definitive.

%% file: content/bias.tex
Machine learning (ML) is becoming increasingly popular in statistics 
courses to help students analyze complex data patterns and create 
predictive models. Learning ML techniques, especially with considerations over social fairness, can help students understand how to build models that not only perform well but also address issues like bias and inequality in real-world applications.

The goal of ``fair'' machine learning is as follows: in 
predicting $Y$ from $X$, we want to minimize the influence of any sensitive variables  $S$ \citep{oneto2020fairness}. In evaluating an application for a mortgage, we hope that our assessment will not discriminate against women or minorities. Such issues have been the subject of intensive research in recent years. There is even a conference dedicated to this effort, the 
\emph{ACM Conference on Fairness, Accountability, and Transparency} 
(ACM FAccT). In this section, we present a brief overview of fairness 
in ML and some of \texttt{dsld}'s capabilities in this regard.

The ``Hello World'' example for this realm of research involves the 
COMPAS (Correctional Offender Management Profiling for Alternative 
Sanctions) algorithm developed by Northpointe, a commercial entity. 
The algorithm was designed to predict a defendant's likelihood of 
recidivism and aid judges in sentencing decisions. Note that, as a 
commercial product, the algorithm is a ``black-box," of completely 
unknown details.

The algorithm faced criticism following an investigation by the 
publication \emph{ProPublica}, which claimed that the algorithm 
exhibited bias against Black defendants \citep{propublica1}. Northpointe contested these findings, arguing that \emph{ProPublica}'s analysis was flawed. \emph{ProPublica} has issued a rejoinder, but in any case, the debate over COMPAS underscores the critical need to address algorithmic fairness.  

\noindent Regarding fairness, there are several major concerns in machine learning:

\begin{itemize}

    \item \textbf{Defining unfairness:} How do we define an algorithm bias/unfairness? Several measures have been proposed.

    \item \textbf{Reducing unfairness:} How can we ameliorate an algorithm's unfairness while still maintaining an acceptable level of accuracy? This is known as the Fairness-Utility tradeoff.

    \item \textbf{Dealing with proxies:} Typically there will be proxy
    variables which, through correlation with \emph{S}, can result in the latter influencing $\widehat{Y}$ even if \emph{S} is omitted from the analysis entirely.

\end{itemize}

\noindent The \texttt{dsld} package provides a convenient and simple 
interface to a range of fairness-constrained modeling approaches.  

We emphasize fairness–accuracy tradeoffs using the \texttt{COMPAS} and \texttt{svcensus} datasets in both regression and classification settings. In the \texttt{COMPAS} data, we predict the probability of recidivism with race as the sensitive feature, while in the \texttt{svcensus} data, we predict wage income with gender and age as the sensitive variables. In both applications, additional care is required to address fairness in sensitive demographics, and we discuss necessary considerations that students should be aware of. Both datasets are included in the \texttt{dsld} package for further exploration and analysis.

\vspace{-1em}
\subsection{Measuring Fairness and Accuracy}

One measure of fairness is to require $\widehat{Y}$ and $S$ to be statistically independent \citep{lum}. However, this approach excludes proxies and makes it difficult to balance a fairness-utility tradeoff. Many other fairness measures have been proposed, but they are often restrictive and not easily widely applicable.  

A broader view is to minimize the correlation between $\widehat{Y}$ and $S$. See, for example, \cite{li2023fairness}, \cite{kozodoi2022fairness}, \cite{Deho2022intro}, \cite{Mary2019FairnessAwareLF}, and \cite{baharlouei2019r}, which build on prior work by \cite{lee2022maximal} and \cite{roh2023improving}.  

\noindent The \texttt{dsld} package continues this theme:

\begin{itemize}

\item \textbf{Measuring Fairness:} Fairness can be measured by the correlation between $\widehat{Y}$ and $S$, with the goal of minimizing the absolute correlation.  

\item \textbf{Measuring Accuracy:} In regression, accuracy is measured by mean absolute prediction error (MAPE). In classification, we use the overall misclassification rate. Additional metrics such as F1 score, sensitivity, and specificity should also be considered.  

\end{itemize} 

\vspace{-1em}
\subsection{Proxy Hunting}

A proxy is a variable that indirectly represents a protected attribute,
potentially introducing bias in decision-making even when the protected
attribute is not explicitly used. For example, race or religion may be
linked to a city or neighborhood in a city \citep{mitopencourseware}. Consider use of \texttt{dsldOHunting()} on the \texttt{svcensus} example:

\begin{lstlisting}[language=, breaklines=true, breakatwhitespace=true]
# only consider "gender" for now
> dsldOHunting(data = svcensus, yName = "wageinc", sName = "gender") 

#               age   educ.14  educ.16  educ.zzzOther  occ.100  occ.101
# gender.female 0.009  -0.011   -0.046        0.031     0.113    0.015
# gender.male  -0.009   0.011    0.046       -0.031    -0.113   -0.015

#               occ.102  occ.106  occ.140  occ.141  wkswrkd
# gender.female  -0.014   0.064   -0.041   -0.139  -0.035
# gender.male     0.014  -0.064    0.041    0.139   0.035
\end{lstlisting}

\noindent 
Based on the correlations above, occupation may act as a proxy for gender, certainly plausible if some engineering fields are more male-dominated. This provides a useful starting point for fairness analysis. 

\vspace{-1em}
\subsection{Relevant Methods Provided by dsld}

The \texttt{dsld} package provides wrappers for several functions from the \texttt{fairML} package \citep{scutari2023fairml} and for the
\texttt{Explicitly Deweighted Features} (EDF) methods developed in
\cite{matloff2022novel}. We first consider the \texttt{fairML} functions.  

\cite{pmlr-v80-komiyama18a}, \cite{zafar2017fairness}, and
\cite{scutari2023fairml} propose ridge-regression-style approaches to
reduce the influence of $S$ on $\widehat{Y}$. These methods are implemented in the \texttt{fairML} package \citep{fairml}, with \texttt{dsld} providing interfaces. The goal here is to directly minimize the effect of $S$; note that proxies are not considered in the \texttt{fairML} functions.  

Let's fit a Fair Generalized Ridge Regression model---a fairness-adapted version of generalized linear models---to predict the probability of recidivism in the \texttt{COMPAS} dataset. The \texttt{unfairness} parameter must lie in the interval $(0,1]$. Values close to 0 prioritize fairness. Assume a train/test split has already been created. 

\begin{lstlisting}[language=, breaklines=true, breakatwhitespace=true]
# train fgrrm model, unfairness = 0.05 -> prioritize fairness
> fgrrmOut <- dsldFgrrm(data = COMPAS_trn, yName = "two_year_recid",  sName = "age", unfairness = 0.05, definition = "sp-komiyama", yesYVal = "Yes")  

# get training data correlations and misclassification rate
> fgrrmOut$trainCorrs
#     "feature"  "correlation"
# 1.     age     -0.06154745

> fgrrmOut$trainAcc     
#  0.2635432
\end{lstlisting}

\noindent To predict on the test set and return testing correlations:  

\begin{lstlisting}[language=, breaklines=true, breakatwhitespace=true]
> preds <- predict(fgrrmOut, COMPAS_tst)  # returns test preds & corrs
> preds$correlations

#    "feature"  "correlation"
# 1     age      -0.1281269
\end{lstlisting}

\noindent Different values of \texttt{unfairness} yield different tradeoffs between fairness and predictive accuracy.  

\noindent The call forms for all \texttt{dsldFairML} functions are as follows:  

\begin{lstlisting}[language=, breaklines=true, breakatwhitespace=true]
> dsldFrrm(data, yName, sName, unfairness, definition = "sp-komiyama", lambda = 0, save.auxiliary = FALSE)
> dsldFgrrm(data, yName, sName, unfairness, definition = "sp-komiyama", family = "binomial", lambda = 0, save.auxiliary = FALSE, yesYVal)
> dsldNclm(data, yName, sName, unfairness, covfun = cov, lambda = 0, save.auxiliary = FALSE)
> dsldZlm(data, yName, sName, unfairness)
> dsldZlrm(data, yName, sName, unfairness, yesYVal)
\end{lstlisting}

The \texttt{dsld} package also includes wrappers for functions based on
EDF \citep{matloff2022novel}. This approach omits $S$ entirely and
allows control over the degree of influence for proxy variables via
user-defined ridge-like hyperparameters. Methods include linear and
logistic regression (with separate ``$\lambda$'' values for each proxy),
random forests (with adjusted probabilities for proxy variables at each
node split), and k-nearest neighbors (k-NN) (using weighted Euclidean
distance to $S$ for proxies).

Let's apply fair k-NN on the \texttt{svcensus} dataset, deweighting the proxy variable \texttt{occupation} to 0.1. 

\begin{lstlisting}[language=, breaklines=true, breakatwhitespace=true]
> knnOut <- dsldQeFairKNN(data = train, yName = "wageinc", sNames = c("gender", "age"), deweightPars = list("occ" = 0.1), k = 25)

# training results
> knnOut$trainAcc
# 26966.77

> knnOut$trainCorrs
#          feature correlation
# 1 gender==female -0.05929727
# 2   gender==male  0.05929727
# 3            age  0.02470862
\end{lstlisting}

\noindent Users can similarly generate predictions on new test data and compute associated correlations.  

\noindent The call forms for all \texttt{dsldQeFairML} functions are as follows:  

\begin{lstlisting}[language=, breaklines=true, breakatwhitespace=true]
> dsldQeFairKNN(data, yName, sNames, deweightPars = NULL, yesYVal = NULL, k = 25, scaleX = TRUE)
> dsldQeFairRF(data, yName, sNames, deweightPars = NULL, nTree = 500, minNodeSize = 10, mtry = floor(sqrt(ncol(data))), yesYVal = NULL)
> dsldQeFairRidgeLin(data, yName, sNames, deweightPars = NULL)
> dsldQeFairRidgeLog(data, yName, sNames, deweightPars = NULL, yesYVal)
\end{lstlisting}

\noindent The \texttt{KNN} and \texttt{RF} functions can be used for either regression or classification settings, depending on the inputs to \texttt{yName} and \texttt{yesYVal}.

\vspace{-1em}
\subsection{The Fairness-Utility Tradeoff}

The fairness--utility trade-off means that prioritizing fairness typically reduces predictive accuracy. The package provides \texttt{dsldFairUtils()}, which conducts k-fold cross-validation with grid searches over different values of the \texttt{unfairness} and \texttt{deweightPars} parameters to evaluate this balance. We show its use with two examples building on the cases above.

Let's begin with the \texttt{COMPAS} example. We test \texttt{dsldFgrrm()} using several values of the \texttt{unfairness}
parameter ranging from 0.01 to 0.99.

\begin{lstlisting}[language=, breaklines=true, breakatwhitespace=true]
> dsldFairUtils(data = compas1, yName = "two_year_recid", sName = "race", dsldFTNName = "dsldFgrrm", unfairness = c(0.9, 0.6, 0.1,0.05, 0.005), deweightPars = NULL, yesYVal = 'Yes', k_folds = 5, model_args = NULL)

  unfair testAcc  Black  White  Other  Hispanic  Asian  Native-American
1  0.900  0.2625 0.2737 -0.176 -0.085  -0.105   -0.067   0.011
2  0.600  0.2625 0.2737 -0.176 -0.085  -0.105   -0.067   0.011
3  0.100  0.2663 0.1652 -0.109 -0.047  -0.060   -0.044   0.010
4  0.050  0.2695 0.0833 -0.057 -0.021  -0.028   -0.025   0.007
5  0.005  0.2721 0.0162 -0.012 -0.001  -0.005   -0.007   0.003
\end{lstlisting}

The output of \texttt{dsldFairUtils()} reports the misclassification rate with correlation values for each race category at specified unfairness levels. The results show the fairness--utility trade-off: as the correlation between $S$ and predicted $Y$ decreases, the overall misclassification rate increases.

Next, we use the \texttt{svcensus} dataset and apply \texttt{dsldQeFairKNN()}, deweighting the ``occupation'' variable over values ranging from 0.01 to 0.99.

\begin{lstlisting}[language=, breaklines=true, breakatwhitespace=true]
> dsldFairUtils(data = svcensus, yName ='wageinc', sName = c('gender', 'age'), dsldFTNName = 'dsldQeFairKNN', k_folds = 5, model_args = list(k = 25), deweightPars = list('occ' = c(0.9,0.8,0.5,0.3,0.1,0.05,0.01)))

    occ  testAcc   male   female    age
1 0.90 25860.33  0.1020 -0.1020 0.0166
2 0.80 26010.03  0.0942 -0.0942 0.0073
3 0.50 26240.20  0.0953 -0.0953 0.0165
4 0.30 29333.87  0.0565 -0.0565 0.0327
5 0.10 28443.07  0.0625 -0.0625 0.0155
6 0.05 27943.57  0.0592 -0.0592 0.0152
7 0.01 29441.34  0.0599 -0.0599 0.0217
\end{lstlisting}

Similar to the \texttt{COMPAS} case, we again note a fairness--utility trade-off. The MAPE increases by more than $4{,}000$ when fairness is prioritized, while the correlation between gender and predicted income is reduced by about half. Since we chose occupation as a proxy for gender, these results align with those from \texttt{dsldOHunting()}.

Fairness in machine learning involves practical considerations for sensitive groups and requires more nuance than only prioritizing accuracy. The fairness–utility trade-off highlights the need to discuss acceptable balances, while the impact of proxies encourages students to carefully consider how feature sets interact with each other. In both classroom settings and applied contexts, students and researchers are encouraged to test different methods across datasets to explore fairness results in practice.


%% file: content/dsldPy.tex
\noindent \texttt{dsldPy} provides Python implementations, built with \texttt{rpy2}, for all 24 functions in the \texttt{dsld} R package. Users must perform some minimal data preprocessing before running these functions. Each Python function is prefixed with \texttt{dsldPy}. This section highlights the high-level interface for interacting with the software in Python.

The first step is to use \texttt{preprocess\_data()} to convert the Python dataset into an R data frame object and assign numeric and categorical variables. The \texttt{svcensus} dataset is used for all examples in this section.  

\begin{lstlisting}[language=, breaklines=true, breakatwhitespace=true]
## data can initially be in either .csv or .rData
> categorical_features = ['educ', 'occ', 'gender']
> numeric_features = ['age', 'wageinc', 'wkswrkd']
> svcensus = preprocess_data(data = svcensus, cat_features = categorical_features, num_features = numeric_features)
\end{lstlisting}

\noindent This R data frame object can be passed directly into all functions. 

\vspace{0.5em}
\noindent \textbf{Graphical Functions}

\noindent All \texttt{dsldPy} functions are called using the same arguments as their R equivalents. For example, \texttt{dsldPyScatterPlot3D()}:  

\begin{lstlisting}[language=, breaklines=true, breakatwhitespace=true]
> dsldPyScatterPlot3D(data = svcensus, yNames = ['wageinc', 'wkswrkd', 'age'], sName = 'gender')
\end{lstlisting}

\begin{figure}[h]
    \centering
    \includegraphics[width=0.70\textwidth]{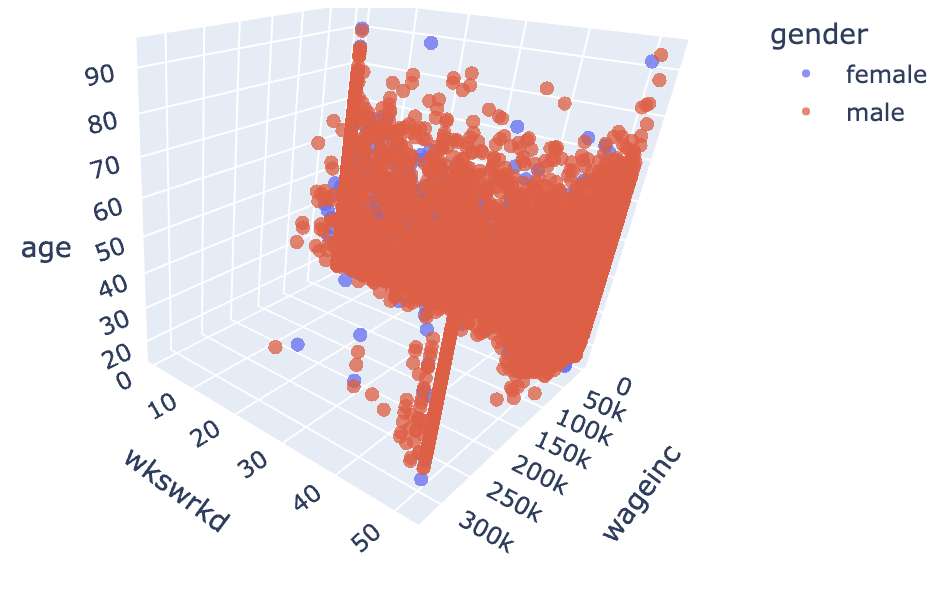}
    \caption{3D scatter plot showing 'wageinc', 'wkswrkd', 'age' segmented by gender.}
    \label{fig:pyScatterPlot}
\end{figure}

Other plotting functions include \texttt{dsldPyConditsDisparity()}, \texttt{dsldPyConfounders()}, \texttt{dsldPyDensityByS()}, \texttt{dsldPyFreqPCoord()}, and \texttt{dsldPyIamb()} with identical arguments to their R versions.  

\vspace{0.5em}
\noindent \textbf{Analytical Function Example}

\noindent Analytical \texttt{dsldPy} functions return Python objects
that can be used in downstream applications. For example,
\texttt{dsldPyLinear()}:

\begin{lstlisting}[language=, breaklines=true, breakatwhitespace=true]
> linOut = dsldPyLinear(data = svcensus, yName = 'wageinc', sName = 'gender', interactions = True)
\end{lstlisting}

\noindent The raw output of \texttt{dsldPyLinear()} is not directly useful. Instead, users can call auxiliary functions to extract necessary results. Note that \texttt{interactions} is set to \texttt{TRUE}.

\begin{lstlisting}[language=, breaklines=true, breakatwhitespace=true]
> dsldPyLinearSummary(linOut)   

{'female': 
           Covariate   Estimate   StandardError     PValue
 1    (Intercept) -11437.040280    2504.918920  4.975159e-06
 2            age    459.798839      44.746473  0.000000e+00
 3         educ16  -1338.140583    3297.967440  6.849285e-01
 4   educzzzOther -11267.385387    1142.379104  0.000000e+00
 5         occ101   -865.939700    1244.706621  4.866181e-01
 6         occ102   9572.851611    1156.552877  2.220446e-16
 7         occ106  -1257.295281    2350.460563  5.927090e-01
 8         occ140   1895.227344    2944.192588  5.197588e-01
 9         occ141   7673.144094    2074.943603  2.173038e-04
 10       wkswrkd   1118.938931      28.442777  0.000000e+00,
 
 'male':  
           Covariate   Estimate   StandardError     PValue
 1    (Intercept) -17143.284561    2141.555691  1.110223e-15
 2            age    473.246288      32.755864  0.000000e+00
 3         educ16   8635.003046    1993.079189  1.474299e-05
 4   educzzzOther -15011.263427     914.294549  0.000000e+00
 5         occ101   3108.187529    1140.756820  6.436550e-03
 6         occ102  15334.830888    1048.633118  0.000000e+00
 7         occ106   2717.392933    2793.202366  3.306232e-01
 8         occ140  14213.551015    1942.295688  2.517986e-13
 9         occ141  12018.178603    1230.058443  0.000000e+00
 10       wkswrkd   1405.045087      26.412856  0.000000e+00,
 
 'Sensitive Factor Level Comparisons': 
    Factors Compared  New Data Row     Estimates  Standard Errors
 1    female - male             1 -10031.389242      2152.555273
 2    female - male             2  -5749.789343      1526.624526
 3    female - male             3 -24856.540762      3849.399483
 4    female - male             4  -6102.013919      1333.947634
 5    female - male             5  -9712.531638      1724.083705}
\end{lstlisting}

The function \texttt{dsldPyLinearSummary()} returns a dictionary of
model summary tables (as pandas DataFrames) for each level of the sensitive variable. The entry \texttt{Sensitive Factor Level Comparisons} reports the differences in predicted estimates between pairs of sensitive levels evaluated at the values specified in \texttt{sComparisonPts} when interactions are \texttt{TRUE}.\footnote{If \texttt{sComparisonPts} is \texttt{NULL}, then 5 rows of original data are randomly selected.}

Additional auxiliary functions are provided: 
\texttt{dsldPyLinearCoef()} returns a dictionary of coefficient arrays by sensitive level; \texttt{dsldPyLinearVCov()} returns a dictionary of
covariance matrices by sensitive level; and \texttt{dsldPyLinearPredict()} returns dataframe of predictions and standard errors.

\begin{lstlisting}[language=, breaklines=true, breakatwhitespace=true]
# newData has two rows (exclude Y and S when interactions are present)
> preds = dsldPyLinearPredict(linOut, newData)
> preds

   level   row  prediction   standardError
1  female    1   68181.51        1035.44
2  female    2   17706.78        1148.63
3    male    1   80047.30         823.97
4    male    2   18505.62        1022.26
\end{lstlisting}

\texttt{dsldPyLinearPredict()} returns individual predictions for each level as in \texttt{interactions} case. This setup also extends naturally to \texttt{dsldPyLogit()} for the classification case. For comparisons of sensitive levels using non-parametric models, \texttt{dsldPyML} returns the test error and predictions on the \texttt{sComparisonPts} data for each sensitive level.  

\begin{lstlisting}[language=, breaklines=true, breakatwhitespace=true]
> mlOut = dsldPyML(data = svcensus, yName = 'wageinc', sName ='gender', qeMLftnName = 'qeKNN', sComparisonPts = 'rand5')
> mlOut

({'testAcc: female': 22311.502857142852, 
  'testAcc: male':   26384.274000000005},
  row       age      educ     occ  wkswrkd   female     male
 8236   31.327554  zzzOther  141     52.0  59912.0  69940.0
 14038  60.989090  zzzOther  141     52.0  54824.0  66588.0
 2772   45.356181  zzzOther  102     40.0  64704.0  66620.0
 13047  28.785540  zzzOther  100     52.0  44352.0  64384.0
 15216  42.916519  zzzOther  101     52.0  49952.0  61284.0)
\end{lstlisting}

\noindent Other analytical \texttt{dsldPy} functions include:  

\begin{itemize}
    \item \texttt{dsldPyTakeALookAround}: returns a pandas DataFrame of feature correlations.
    \item \texttt{dsldPyCHunting}: returns a dictionary of importance values for \texttt{Y} and \texttt{S} across features. 
    \item \texttt{dsldPyOHunting}: returns a pandas DataFrame of correlations between \texttt{S} and other features. 
    \item \texttt{dsldPyFrequencyByS}: returns a pandas DataFrame of frequency tables for sensitive variables and categorical features.
    \item \texttt{dsldPyMatchedAte}: returns a dictionary with summary output from \texttt{dsldMatchedATE}. 
\end{itemize}

\noindent \textbf{Fair Machine Learning Example}

\noindent Similar to the results above, \texttt{dsldPy} machine learning functions return Python objects that can be used for downstream tasks. Consider the example of \texttt{dsldPyQeFairKNN()}:  

\begin{lstlisting}[language=, breaklines=true, breakatwhitespace=true]
> knnOut = dsldPyQeFairKNN(data = svcensus_train, yName = 'wageinc', sName = 'gender', deweightPars = {'occ': 0.01})

# Print training results
print("train predictions:", knnOut["train_predictions"])
print("train accuracy:", knnOut["train_accuracy"])
print("train correlations:", knnOut["train_correlations"])

train predictions: [72300.94, 60485.95, 54039.50, 73667.12, ... 2719.56]
train accuracy: 25449.5408
train correlations: [("gender==female", -0.041989047481350954),
                     ("gender==male",    0.041989047481350954)]
\end{lstlisting}

\noindent Prediction on test data is also straightforward and presents correlations with the sensitive variable:  

\begin{lstlisting}[language=, breaklines=true, breakatwhitespace=true]
### predict() on test data
> preds = dsldPyFairML_Predict(knnOut, svcensus_test)

# print test predictions, correlations
> print(f"test predictions: {preds['test_predictions']}")
> print(f"test correlations: {preds['test_correlations']}")

test predictions: [72389.22, 57325.39, 83349.46, 59201.27, ... 63169.04]
test correlations: [("gender==female", -0.04230130709813465),
                     ("gender==male",    0.042301307098134656)]
\end{lstlisting}

To run k-fold cross validation and evaluate multiple points along the fairness–utility tradeoff, use \texttt{dsldPyFairUtils()}, which returns a pandas data frame of test errors and correlations for each S-level:  

\begin{lstlisting}[language=, breaklines=true, breakatwhitespace=true]
> dsldPyFairUtils(data = svcensus_train, yName = 'wageinc', sName = 'gender', dsldFTNname = "dsldQeFairKNN", deweightPars = {'occ': [0.9 ,0.8 ,0.5 ,0.3 ,0.1 ,0.05 ,0.01]}, k_folds = 10)

     occ      testAcc	  gender==female	gender==male
1	0.90	25983.465304	-0.063162	      0.063162
2	0.80	25959.389920	-0.063797	      0.063797
3	0.50	25845.586591	-0.064265	      0.064265
4	0.30	27451.434721	-0.034131	      0.034131
5	0.10	26980.586397	-0.028072	      0.028072
6	0.05	26624.466493	-0.033655	      0.033655
7	0.01	26217.305465	-0.039768	      0.039768
\end{lstlisting}

These interfaces make it straightforward to use the \texttt{dsld} package in Python with full equivalence to the R functions. They allow students and researchers to apply the methods effectively and with minimal overhead for various statistics and machine learning applications.

%% file: content/discussion.tex
In this paper, we introduce ``Data Science Looks at Discrimination" 
(\texttt{dsld}) as a powerful tool for statistical education and 
machine learning through the lens of discrimination analysis and 
fair machine learning. The software includes analytical and graphical 
tools for detailed exploratory data analysis, allowing students to 
explore and visualize potential sources of bias. Additionally, the 
fair machine learning wrappers provide several fairness-constrained 
methods for easy deployment of fair ML algorithms. The accompanying 
Quarto book offers a comprehensive introduction to key statistical 
principles, using real-world examples to demonstrate \texttt{dsld} methods. It is also accessible to students with just high school mathematics, equipping them with the knowledge and tools to conduct data analysis and apply machine learning systems responsibly.

Instructors and students are both encouraged to apply these methods to 
their own datasets for further analysis. The methods provided by the 
\texttt{dsld} package serve as a valuable addition to introductory 
statistics courses, helping students connect theoretical statistical 
concepts to practical use cases.